\documentstyle[prc,aps,epsf,epsfig,amssymb,amsbsy,amstext,amsfonts]{revtex}
\begin{document}
\draft
\author{M.J. Vicente Vacas and E. Oset}
\address{Departamento de F\'{\i}sica Te\'orica and IFIC, \\
Centro Mixto Universidad de Valencia-CSIC, \\
Institutos de Investigaci\'on de Paterna, Apdo. correos 2085,\\
46071, Valencia, Spain}
\date{\today}
\title{Nuclear medium effects on the $\sigma$ mass and width}
\maketitle
\begin{abstract}
The meson meson interaction in the scalar isoscalar channel at finite baryonic
density is studied in the framework of  a chiral
unitary approach which generates the $f_0$ and $\sigma$ resonances and
reproduces well the meson meson phase shifts in vacuum. We investigate the
$\sigma$ spectral function, its mass and its width as a function of the 
baryon density and discuss possible experimental signatures in the 
$(\pi,2\pi)$ and the $(\gamma,2\pi)$ reactions.
\end{abstract}

\pacs{14.40.Aq; 14.40.Cs; 13.75.Lb}

\section{Introduction}
In the past few years  much effort has been devoted to the study of 
the $\pi\pi$ scattering amplitude in nuclei and its possible
experimental signals, like  the enhancement of the $\pi\pi$ invariant 
mass distributions close to the two pion threshold, seen in the experiments 
of pion induced two pion production in nuclei 
\cite{Bonutti:1996ij,Bonutti:1998zw,Camerini:1993ac,bonutti} and the 
large strength shift found in the $\pi^0\pi^0$ channel \cite{Starostin:2000cb}.
 
The idea of strong threshold effects due to the $\pi \pi$ interaction in a
dense nuclear medium was first suggested in ref. \cite{Schuck:1988jn}.  These
effects would show up  in the scalar isoscalar channel and could be interpreted
as a drop of the  sigma mass in  nuclear matter. Later works, paying
attention to  the chiral constraints of the $\pi\pi$ amplitude and trying to
use realistic potentials,   found indeed an appreciable enhancement of the $\pi
\pi$ scattering amplitude  close to threshold 
\cite{Rapp:1996ir,Aouissat:1995sx,Chiang:1998di,Vicente-Vacas:1998vr}.

These theoretical  $\pi\pi$ scattering amplitudes in the nuclear medium 
have been used in the analysis of the $(\pi,\pi\pi)$ reactions in nuclei
\cite{Rapp:1999fx,VicenteVacas:1999xx,VicenteVacas:2000qy}.
Although the modifications of the $\pi \pi$ invariant mass 
distributions produced by the $\pi\pi$ scattering go in the right direction,
the calculations have not been successful in reproducing the full strength
of the medium effects found in the experiment.

In this paper, we will discuss first our approach to the study of the
pion pion interaction in nuclear matter. Then, we will show some results
on the $\sigma$ meson mass and width. Finally, we will discuss the
possibility of finding a clear experimental signature of these
effects in the $(\pi,\pi\pi)$ and the  $(\gamma,\pi\pi)$ reactions.
 
\section{$\pi \pi$ interaction in vacuum}

In this section we  briefly sketch the simple approach of ref.
\cite{Oller:1997ti}, 
which is enough for our purposes. More elaborate studies can be found in 
\cite{Oller:1998ng,Oller:1999hw,Nieves:2000bx}. The basic idea is to solve a 
Bethe Salpeter (BS) equation, which guarantees unitarity, matching the low
energy results to the chiral perturbation theory ($\chi PT$) predictions.
We consider two possible intermediate states, $\pi \pi$ and $K \bar{K}$,  
labelled 1 and 2 in the formulas. The isospin 0 channel is given by the 
following combinations

\begin{eqnarray}
| K \bar{K} > = - \frac{1}{\sqrt{2}} 
| K^+ (\vec{q}) K^- (- \vec{q}) + K^0 (\vec{q}) \bar{K}^0 (- \vec{q})>,\\[2ex]
| \pi \pi >  = - \frac{1}{\sqrt{6}}
| \pi ^+ (\vec{q}) \pi^- (- \vec{q}) 
+ \pi^- (\vec{q}) \pi^+  (- \vec{q}) + \pi^0 (\vec{q})  \pi^0 (- \vec{q})>,
\end{eqnarray}
where $\vec{q}$ is the mesons momentum  in the center of mass (CM)
system. 
  We neglect the $\eta \eta $ channel which is not relevant
at the low energies we are interested in. 
 
\begin{figure}
 \epsfig {figure=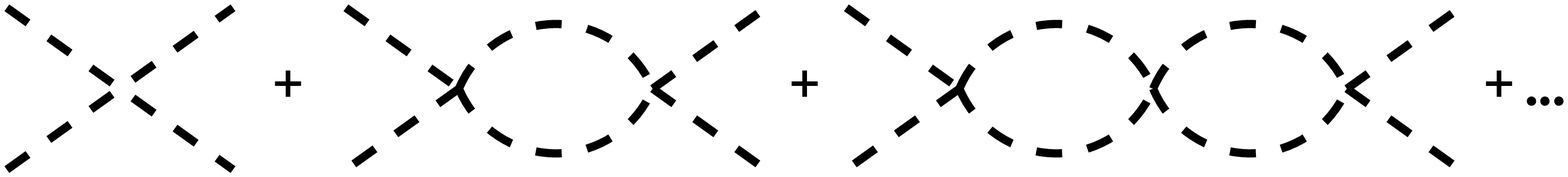,width=14cm}
 \caption{Diagrammatic representation of the Bethe-Salpeter equation.}
 \label{fig:BSF}
\end{figure}

    The BS equation is given by 
\begin{equation}
\label{eq:BS}
T=VGT.
\end{equation}   
Eq. \ref{eq:BS}  is an integral equation and the term $VGT$ 
involves the two mesons one loop divergent integral, (see Fig.~\ref{fig:BSF})  
where  $V$ and $T$ appear off shell. However, for this channel both functions 
can be factorized on shell out of the integral. The remaining off shell part 
can be absorbed by a renormalization of  the coupling constants
as it was shown in refs.\cite{Oller:1997ti,Nieves:1999hp}. 
 Thus, the BS equation becomes purely algebraic and the 
$VGT$ originally inside the loop integral becomes then the product 
of $V$, $G$ and $T$, with $V$ and $T$ the on shell amplitudes, and   
$G$  given by  the  expression  
\begin{equation}
G_{ii} = i \int \frac{d^4 q}{(2 \pi)^4}
\frac{1}{q^2 - m_{1i}^2 + i \epsilon} \; \; 
\frac{1}{(P - q)^2 - m_{2i}^2 + i \epsilon}
\end{equation}
where $P$ is the total momentum of the meson-meson system. This 
integral is regularized with a cut-off adjusted to optimize the fit
to the $\pi-\pi$ phase shifts ($\Lambda=1.03$ GeV).
Finally the potential $V$ appearing in the BS equation is taken from 
the lowest order chiral Lagrangians 

\begin{equation}
L_2 = \frac{1}{12 f^2} < (\partial_\mu \Phi \Phi - \Phi \partial_\mu
\Phi)^2 + M  \Phi^4 \, >
\end{equation}

\noindent
where the symbol $< >$ indicates the trace in flavour space,
$f$ is the pion decay constant and $\Phi$, $M$ are the meson and mass $SU (3)$ 
matrices.

In s-wave and for isospin zero the only pieces needed in our calculation
are given by
\begin{eqnarray}
V_{11} = - < K \bar{K} | {\cal L}_2 | K \bar{K} > = - \frac{1}{4 f^2}
(3 s + 4 m_K^2 - \sum_i p_i^2)\\[2ex]
V_{21} = - < \pi \pi | {\cal L}_2 | K \bar{K} > = - \frac{1}{3 \sqrt{12} f^2}
( \frac{9}{2}s + 3 m_K^2 + 3m_\pi^2 - \frac{3}{2}  \sum_i p_i^2)\\[2ex]
V_{22} = - < \pi \pi | {\cal L}_2 | \pi \pi  > = - \frac{1}{9  f^2}
(9 s + \frac{15  m_\pi^2}{2}  - 3 \sum_i p_i^2)
\end{eqnarray}

This model reproduces well phase shifts and inelasticities up to about 
1.2 GeV. The $\sigma$ and $f_0 (980)$ resonances appear as poles of the 
scattering amplitude. The coupling of channels is essential to produce the 
$f_0 (980)$ resonance, while the $\sigma$ pole is little
affected by the coupling of the pions to $K \bar{K}$
\cite{Oller:1997ti}. 

\section{$\pi \pi$ scattering in the nuclear medium.}

As we are mainly interested in the low energy region, which is not very
sensitive to the kaon channels, we will only consider the nuclear medium 
effects on the pions. The main changes of the pion propagation in the 
nuclear medium come from the p-wave selfenergy, produced basically 
by the the coupling of pions to particle-hole ($ph$) and Delta-hole 
($\Delta h$) excitations \cite{Oset:1990ey}. 

\begin{figure}[htb]
 \begin{center}
\epsfig{height=2.2cm,width=12.2cm,angle=0, figure=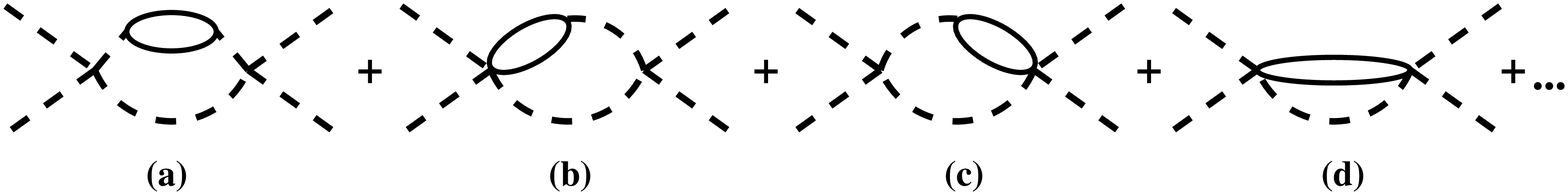}
\caption{Terms of the meson-meson scattering amplitude accounting for
$ph$ and $\Delta h$ excitation.}
 \label{fig:BSF2}
 \end{center}
\end{figure}

Thus, the new BS equation now will include the 
diagrams of Fig. \ref{fig:BSF2} where the solid line bubbles represent
the $ph$ and $\Delta h$ excitations.

In fact, as it was shown in \cite{Chanfray:1999nn}, the contact terms with
the $ph$ ($\Delta h$) excitations of diagrams (b)(c)(d) cancel the 
off shell contribution  from the meson meson vertices in the term of Fig.
\ref{fig:BSF2}(a). 
Hence, we just need to calculate the diagrams of the free type 
(Fig. \ref{fig:BSF}) and those of Fig. \ref{fig:BSF2}(a). 

\begin{figure}[htb]
 \begin{center}
\epsfig{height=2.5cm,width=5.cm,angle=0,figure=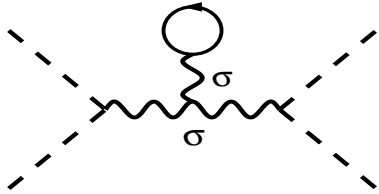}
 \caption{ Tadpole $\sigma$ selfenergy diagram.}
 \label{fig:SIG}
 \end{center}
\end{figure}

There are other medium corrections, like the one 
 depicted in Fig. \ref{fig:SIG}
which has been considered in several papers like 
\cite{Hatsuda:1999kd,Kunihiro:1999mt,Jido:2001bw} using the linear sigma model. 
The coupling of the sigma to the nucleons, can be borrowed from the Bonn  
potential \cite{Machleidt:1987hj}. 

\begin{figure}[htb]
 \begin{center}
\epsfig{height=2.0cm,width=12.5cm,angle=0,figure=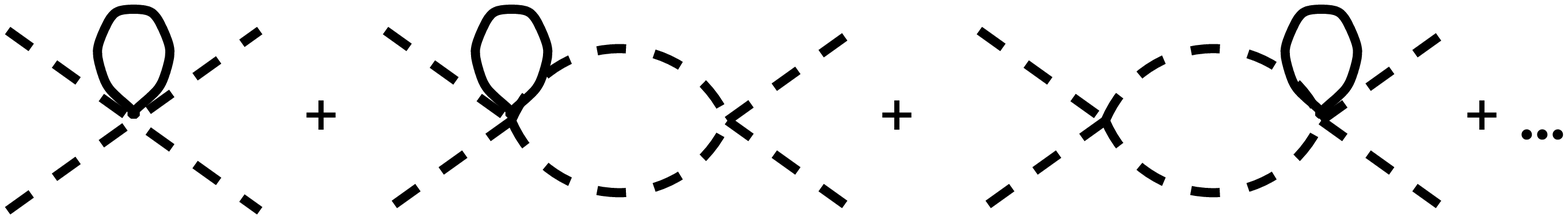}
 \caption{ Some tadpole diagrams contributing to the meson-meson scattering 
 amplitude }
 \label{fig:SIGPT}
 \end{center}
\end{figure}

In our approach, we start with only pseudoscalar meson fields. The sigma is 
generated dynamically  through the rescattering of the pions. A possible   
analog to the diagram of Fig. \ref{fig:SIG}  is given by the diagrams of 
Fig. \ref{fig:SIGPT}.
 In order to evaluate the contribution of these diagrams we use the
chiral Lagrangians involving the octet of baryons and the octet of 
pseudoscalar mesons and these  terms result to be  proportional to 
$\bar{p}\gamma^\mu p-\bar{n}\gamma^\mu n$ and thus vanish in 
symmetric nuclear matter.

\begin{figure}[htb]
 \begin{center}
\epsfig{height=3.0cm,width=5.5cm,angle=0,figure=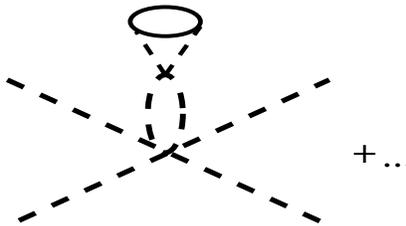}
 \caption{Higher order tadpole diagram. }
 \label{fig:SIGPT2}
 \end{center}
\end{figure}

Also we would need to consider diagrams of the type shown in Fig. \ref{fig:SIGPT2}
We also get a negligible contribution to the the $\pi \pi$
scattering in the nuclear medium from this diagram \cite{Oset:2000ev}. 

\begin{figure}[htb]
 \begin{center}
\epsfig{height=3.0cm,width=5.5cm,angle=0,figure=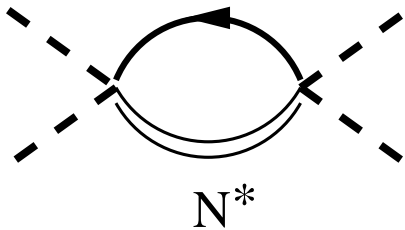}
 \caption{First order resonance-hole contribution to the meson meson 
 scattering in the scalar isoscalar channel.}
 \label{fig:ROP1}
 \end{center}
\end{figure}

Finally, we could also include  excitation of resonances from the occupied 
nucleon states by the pair of mesons. The major contribution comes from 
the $N^*(1440)$ Roper resonance which couples to two pions in the scalar 
isoscalar channel and  plays an important role in the  
$\pi N \rightarrow \pi \pi N$ reaction 
\cite{Oset:1985wt,Sossi:1993zw} and other two pion production processes
like  the $N N \rightarrow N N  \pi\pi $ reaction close to threshold 
\cite{Alvarez-Ruso:1998mx,Alvarez-Ruso:1999xg}. 
 The lowest order Roper contribution diagram is shown in Fig. \ref{fig:ROP1}.
Details on the inclusion of this mechanism, the determination of the coupling
constants and its effects, which are also small, can be found in ref. 
\cite{Oset:2000ev}.

\begin{figure}[htb]
 \begin{center}
\epsfig{height=18.cm,width=16.cm,angle=0,figure=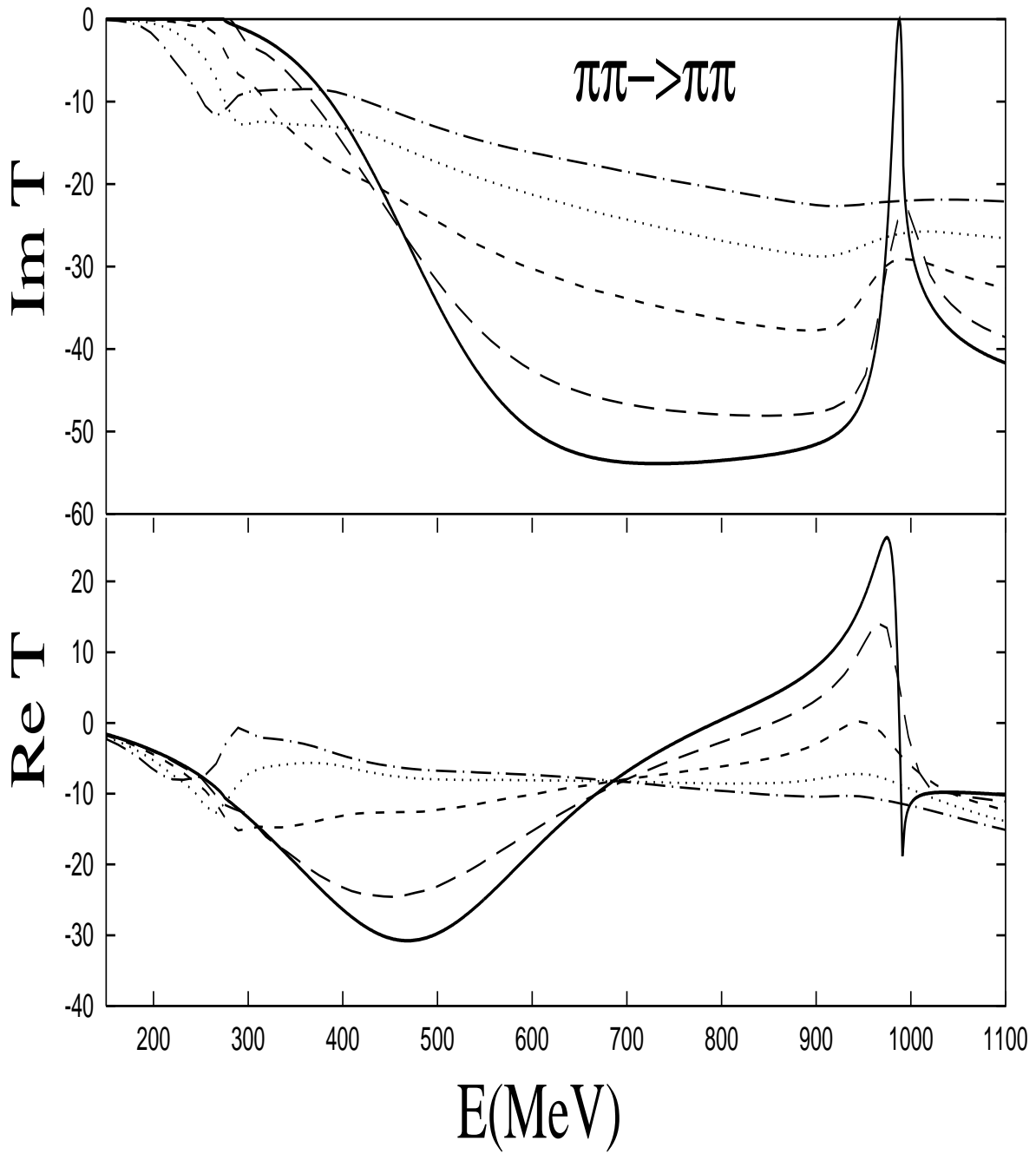}
 \caption{Real and imaginary part of the pion-pion scattering amplitude 
in the S=I=0 channel as a function of the center of mass energy for different 
nuclear densities.Solid line, free amplitude; long dashed line, 
$\rho=\rho_0/8$; short dashed line, $\rho=\rho_0/2$; dotted line, 
$\rho=\rho_0$; dashed dotted line, $\rho=1.5\rho_0$.}
 \label{fig:R4}
 \end{center}
\end{figure}

In Fig. \ref{fig:R4} we show our results for the $\pi\pi$ scattering amplitude
as a function of the nuclear density in a range of 
energies from 150 MeV to 1100 MeV.  As one can see in the figure, there is an
accumulation of strength in the imaginary part of the amplitude close to 
the two pion threshold as it was predicted by
 \cite{Schuck:1988jn} and had been  also found in other calculations 
\cite{Aouissat:1995sx}. The possible relation of this  strength to the enhanced
two pion invariant mass distribution in $(\pi,2\pi)$ reactions in nuclei 
at small invariant masses will be discussed in the next section.
The strength below the two pion threshold corresponds to the decay into 
particle-hole states.
At intermediate invariant masses we can see a quite strong
decrease of the strength of the imaginary part, and also large distortions of
the real part of the scattering amplitude. This could  lead to some
effects, not yet investigated, for two pion production reactions.
Below 900 MeV, the effect of the two kaon channel are quite small, and a single 
channel calculation would lead to quite similar results.

 Also the  $f_0(980)$ resonance properties are modified in the medium.
  The position of the resonance does not change much as a function of the 
density.  However, the width increases gradually until the resonant shape
practically disappears above the normal nuclear density.
 
\begin{figure}[htb]
 \begin{center}
\epsfig{height=18.cm,width=16.cm,angle=0,figure=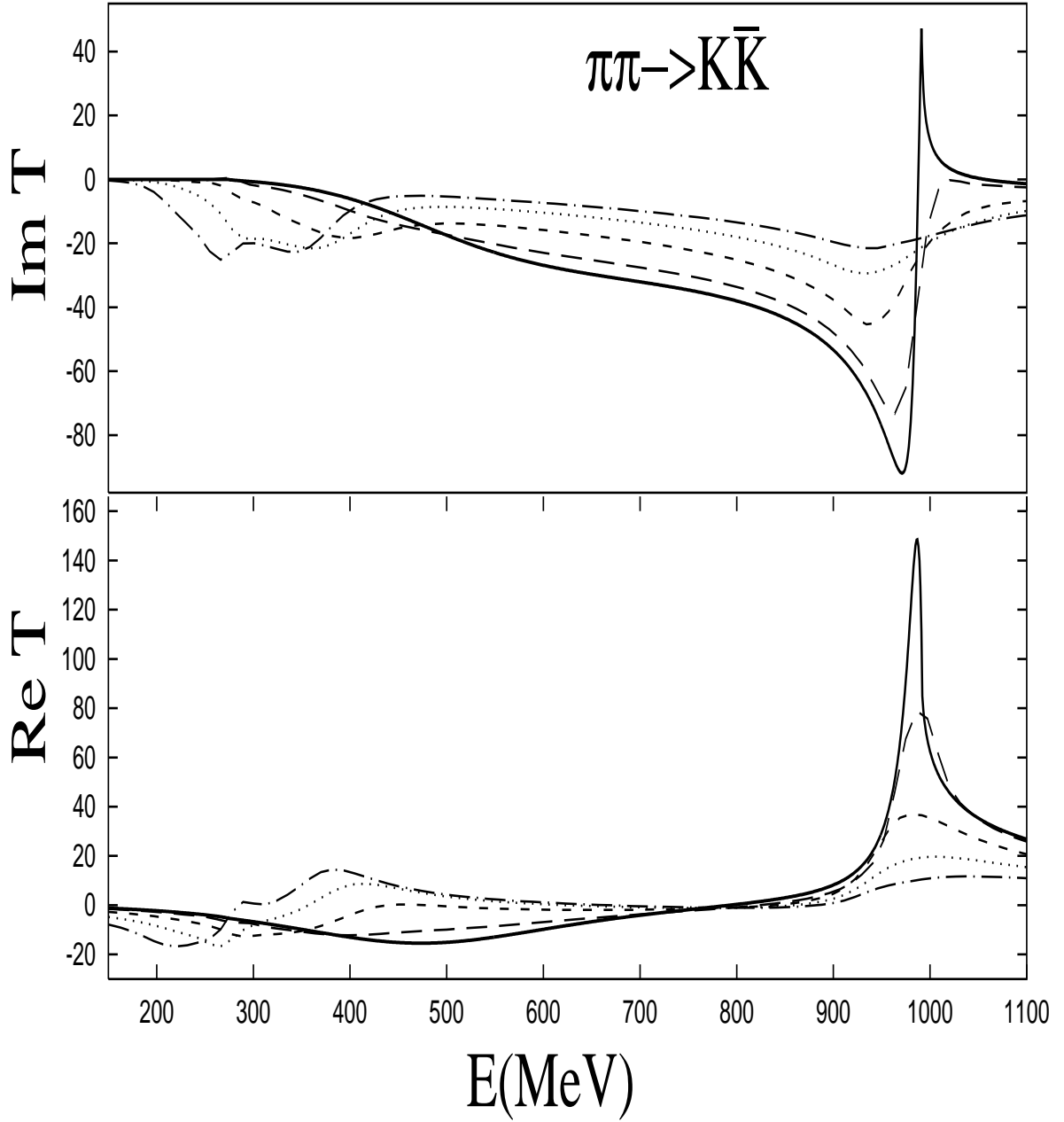}
 \caption{Real and imaginary part of the $\pi\pi\to K\bar{K}$ scattering 
amplitude in the I=0 channel as a function of the center of mass energy 
for different nuclear densities. Meaning of the lines as in Fig. 7}
 \label{fig:R6}
 \end{center}
\end{figure}
  
In Fig. \ref{fig:R6} we show the $\pi \pi\to K \bar{K}$ amplitude. In this 
channel the $f_0(980)$ appears as a Breit Wigner contribution rotated 90 
degrees and it is more easily visible,
due to the absence of a large background of the $\pi \pi$ elastic scattering.
 The resonant-like behavior found around 300 MeV is mainly produced by the 
strong coupling of the two kaon channel to the Roper-hole excitation. 
A detailed discussion on that coupling and results for the 
$K\bar{K}\to K \bar{K}$ amplitude can be found in ref. \cite{Oset:2000ev}.

\section{$\sigma$ meson mass and width in the nuclear medium} 

In this section we will consider the ligthest pole position of the
$\pi\pi$ scattering amplitude in the S=I=0 channel in the complex energy 
plane, which corresponds to the $\sigma$ resonance. In vacuum, the pole 
position occurs at a mass around 470 MeV and the width is around 400 MeV 
\cite{Oller:1997ti}. A compilation of values for these magnitudes obtained in several modern 
analyses can be found in ref. \cite{Markushin:2001kx}. See also Fig. 1 
of ref. \cite{Tornqvist:2000jt}. 

\begin{figure}[htb]
 \begin{center}
\psfig{height=3cm,figure=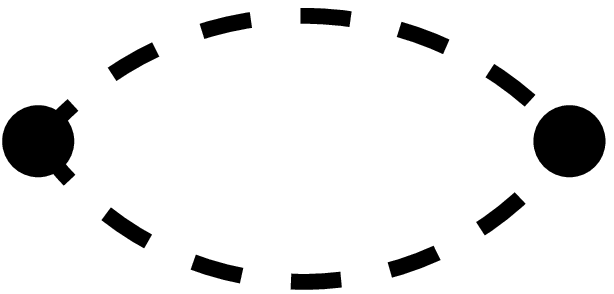}
\caption{}
  \label{fig:R7}
 \end{center}
\end{figure}

\begin{figure}[htb]
 \begin{center}
\psfig{height=12.4cm,angle=-90,figure=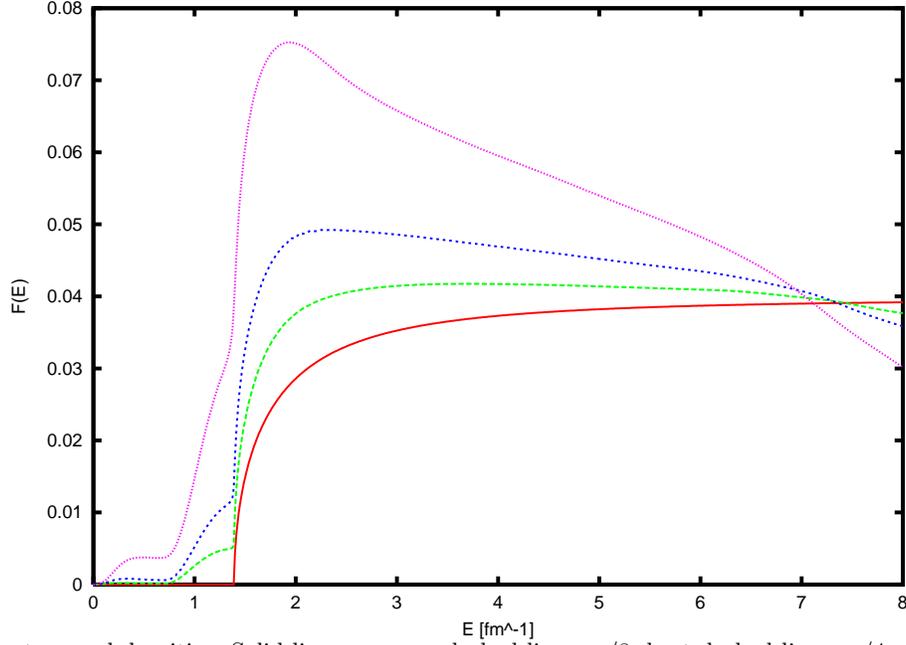}
\caption{$F(E)$ at several densities. Solid line: vacuum, dashed line: 
$\rho_0/8$,short-dashed line: $\rho_0/4$, dotted line: $\rho_0/2$ }
  \label{fig:R8}
 \end{center}
\end{figure}

\begin{figure}[htb]
 \begin{center}
\psfig{height=3.cm,figure=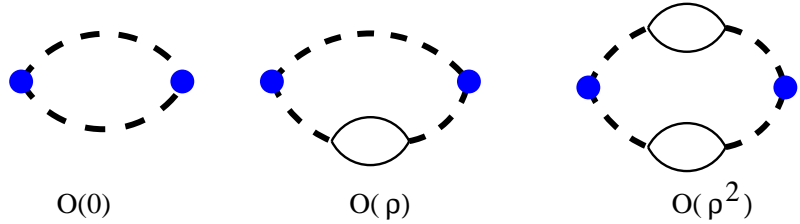}
\caption{$\sigma$ decay channels classified according to their density
dependence}
  \label{fig:R9}
 \end{center}
\end{figure}

The analytical structure of the meson meson scattering amplitude is driven 
by the meson loops depicted in Fig. \ref{fig:R7}, 

which are given by the formula
\begin{equation}
  G_{\pi \pi }\sim \int ^{\infty }_{0}
 d^{4}q\, D_{\pi }(q^{0},\overrightarrow{q})D_{\pi }
 (\sqrt{s}-q^{0},-\overrightarrow{q}) 
\end{equation}
where $D_{\pi }$ is the $\pi$ propagator which, in the nuclei, will incorporate 
the p-wave coupling of the pions to particle-hole and $\Delta$-hole 
excitations.
We can simplify the calculation by using the Lehmann representation for the 
meson propagators
\begin{equation}
 D_{\pi }(q^{0},\overrightarrow{q})=-\frac{2}{\pi }\int ^{\infty }_{0}
dx\, \frac{x\, \, Im\, D_{\pi }(x,\overrightarrow{q})}
{(q^{0})^{2}-x^{2}+i\epsilon }
\end{equation}
After some algebraic manipulation we obtain
\begin{equation}
\label{eq:poles}
 G=\int ^{\infty }_{0}\frac{dE}{2\pi }
(\frac{1}{P^{0}-E+i\epsilon }-\frac{1}{P^{0}+E-i\epsilon })\, F(E) 
\end{equation}
where $F(E)$ is a real function given by
\begin{equation}
 F(E)=\int d^{3}q\, \int ^{E}_{-E}dx\: Im\, D_{\pi }((E+x/2),
\overrightarrow{q})\: Im\, D_{\pi }((E-x/2),\overrightarrow{q}) 
\end{equation}
which in vacuum takes the simple form
\begin{equation}
F_{vacuum}(E)=\frac{p_{\pi }}{4\pi \, E} 
\end{equation}
where 
\begin{equation}
p_{\pi }=\sqrt{\frac{E^{2}}{4}-m_{\pi }^{2}} 
\end{equation}
The function $F(E)$ includes all the phenomenological information on the
pion selfenergy in the nuclear medium. The results for several densities are
shown in Fig.\ref{fig:R8}
The $F(E)$ strength is related to the imaginary part of the $\sigma$ 
propagator and therefore reflects  the energy and density dependence of the
different $\sigma$ decay channels. We can start classifying them according 
to their density dependence (See. Fig.\ref{fig:R9}).
At low densities, the $\sigma$ meson can only decay into two pions,
 $\sigma \rightarrow \pi+\pi$. Therefore we have a threshold at $E=2m_\pi$.
As the density grows the  $\sigma \rightarrow \pi+(Nh)$ decay channel 
becomes relevant and we have a new threshold at $E=m_\pi$, which is clearly
visible in the curve corresponding to  $\rho=\rho_0/8$. Finally, at larger
densities, mechanisms such as $\sigma \rightarrow 2(Nh)$
which go like $\rho^2$  become important. They are possible even at very low
energies.

\begin{figure}[htb]
 \begin{center}
\psfig{height=12.4cm,angle=-90,figure=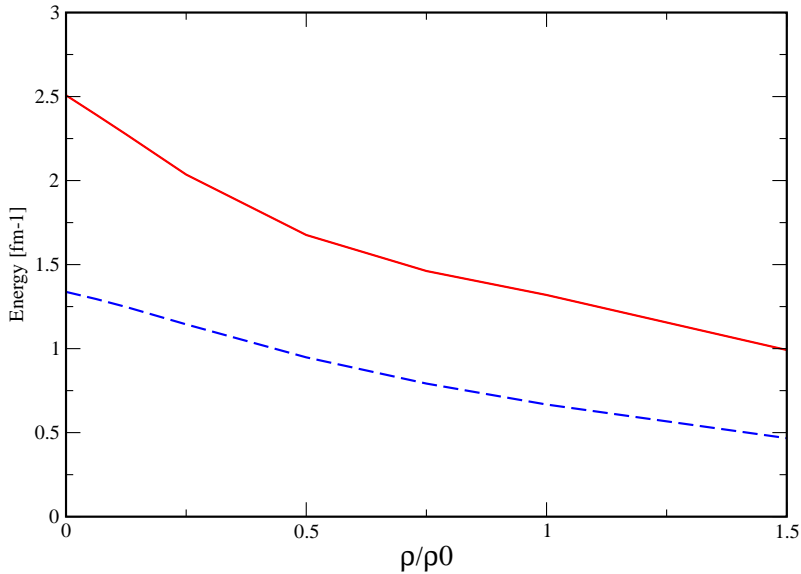}
\caption{$\sigma$ mass (solid line) and half width (dashed line) as a function
of the density}
  \label{fig:R10}
 \end{center}
\end{figure}

Using the two meson propagator of Eq. \ref{eq:poles} we can calculate the 
scattering amplitude and look for the poles in the complex energy plane. The 
results are shown in Fig. \ref{fig:R10} for densities up to 
1.5 $\rho_0$. Note, however, that the calculation is more reliable at low
densities because some contributions of order $\rho^2$ or higher are 
missing.  We find that both mass and width decrease as the density increases,
reaching a mass around 200 MeV and a similar width at 1.5 times the nuclear
density.  
Similar results are found in other models. See for instance Fig. 10 of 
ref. \cite{Hatsuda:2001da}. But we should stress that there the changes
respond to a reduction of the $f_\pi$ value which we have kept constant.
The basic ingredient that drives the mass decrease in our calculation
is the p-wave interaction of the pion with the baryons in the medium.
The change of $f_\pi$ as a fuction of density in \cite{Hatsuda:2001da}
and related works is linked to the renormalization of the axial current in
nuclei, but in standard many body theory these currents are renormalized 
using the same lagrangians that in the evaluation of the pion selfenergy
\cite{rho,Gil:1995ph,Oset:2001iy}. Therefore one should not modify  $f_\pi$ in these latter approaches
to avoid doublecounting.
 In fact a large change at normal nuclear density would be quite
difficult to accommodate to the quite well known pion nucleus phenomenology
(pionic atoms, pion absorption, etc.).

\section{Experimental signatures}

Reactions with two low energy  pions in the final state are the obvious
candidates to test the $\sigma$ meson properties in the nuclear medium. 
Both CHAOS and CB data on the $A(\pi ,\pi +\pi)X$ reaction have been 
extensively discussed in several talks at this conference. Therefore, 
we will just comment briefly a few points. 

In both cases there are clear and strong medium effects in the scalar 
isoscalar channels.  The CHAOS collaboration has  measured several
charge channels that show these effects to be absent in other channels.

 So far, the theoretical attempts have failed to reproduce existing data.
The enhancement obtained at low "$\sigma$" masses by the calculations
is too small.
There are claims that additional strength would be gained once the effects
of the partial restoration of the chiral symmetry are included to the
"standard" nuclear medium effects of the current theoretical analyses.
However, even large changes in the pion pion interaction would affect little
the calculations. The main reason for this being the very large distortion
(quasielastic and absorption) of the initial pions at the CHAOS energies.
Because of this, the reaction happens at the nuclear surface, with a quite 
low effective density.

On the other hand, the $(\pi,\pi^+ + \pi^-)$ and $(\pi, \pi^0 + \pi^0)$
channels are very small at low invariant masses of the two pions system.
Comparing these invariant mass distributions with phase space, and studying 
the microscopic model for the reaction it is clear that there are strong 
cancellations between different mechanisms that contribute to this reaction.
These cancellations could be modified in nuclei  if the relative phases
of the involved mechanisms change even slightly and that could produce a
large part of the observed effects. 

Further and better studies are clearly needed. However, 
$(\pi,\pi\pi)$ reactions will never be free from the two problems pointed out,
the strong initial pion distortion (and therefore small effective density)
and the complicated microscopical mechanisms involved. The $(\gamma,\pi\pi)$ 
process is, at least, free from the first of those problems and look clearly
more promising\cite{Oset:2001iy}. Some preliminary results have already been 
presented at this conference by Volker Metag and could be more conclusive in 
determining the strength of the $\pi-\pi$ interaction in nuclei.

\acknowledgements{This work is partly supproted by DGICYT contract no. 
BFM2000-1326 and by the EEC-TMR Program, EURODAPHNE, Contract No. 
ERBFMRX-CT98-0169.}


\end{document}